\documentclass[aps,prd,12pt,preprint,tightenlines,superscriptaddress,
amsfonts,amssymb,amsmath,byrevtex,showpacs,nofootinbib]{revtex4}
\usepackage{graphics}
\usepackage{epsfig,subfigure}
\newcommand{\be}{\begin{equation}}
\newcommand{\ee}{\end{equation}}
\newcommand{\bea}{\begin{eqnarray}}

\newcommand{\eea}{\end{eqnarray}}

\newcommand{\dR}{\mathbb R}

\begin{document}

\title{Bianchi IX model: Reducing phase space}

\author{Ewa Czuchry}
\affiliation{ Department of Fundamental Research, National Centre
for Nuclear Research, Ho{\.z}a 69, PL-00-681 Warsaw, Poland}
\author{W{\l}odzimierz Piechocki}
\affiliation{ Department of Fundamental Research, National Centre
for Nuclear Research, Ho{\.z}a 69, PL-00-681 Warsaw, Poland}

\date{\today}

\begin{abstract}

The mathematical structure of higher-dimensional physical phase
spaces of the nondiagonal Bianchi IX model is analyzed in the
neighborhood of the cosmological singularity by using dynamical
system methods. Critical points of the Hamiltonian equations
appear at infinities and are of a nonhyperbolic type, which is a
generic feature of the considered singular dynamics.  The
reduction of the kinematical symplectic 2-form to the constraint
surface enables the determination of the physical Hamiltonian.
This procedure lowers the dimensionality of the dynamics arena.
The presented analysis of the phase space is based on canonical
transformations. We test our method for the specific subspace of
the physical phase space. The obtained results encourage further
examination of the dynamics within our approach.

\end{abstract}

\pacs{04.20.-q, 05.45.-a}

\maketitle

\section{Introduction}

The available data of observational cosmology indicate that the
Universe emerged  from a state with extremely high energy
densities of matter fields. Theoretical cosmology,  particularly
the Belinskii-Khalatnikov-Lifshitz (BKL) scenario  (which is
thought to be a generic solution to the Einstein equations near
spacelike singularities), characterizes the spacetime in the
region of the initial cosmological {\it singularity} with
diverging gravitational and matter field invariants
\cite{BKL22,BKL33,Belinski:2009wj}. It has been proved that
symmetries of spacetime such as isotropy and homogeneity are
dynamically {\it unstable} in the evolution towards the
singularity \cite{BKL22,BKL33}. The existence of the {\it general}
solution to the Einstein equations with a cosmological singularity
means that the classical theory is {\it incomplete}.

The best prototype for the BKL scenario is the nondiagonal Bianchi
IX model \cite{BKL33,BKL2}. We expect that obtaining a {\it
quantum} Bianchi IX model may enable  quantization of  the BKL
scenario. Finding the {\it nonsingular} quantum BKL theory would
mean solving, to some extent,  the generic cosmological
singularity problem. Such a quantum theory could be used as a
realistic model of the very early Universe.

In this paper we examine the mathematical {\it structure} of the
physical phase space of the nondiagonal Bianchi IX model near the
cosmological singularity. We are aware of great complexities
specific to the dynamics of the homogeneous models. There exists
comprehensive literature in the field:    textbooks (see, e.g.,
\cite{Wain,Bogo,Ryan}) and review articles (see, e.g.,
\cite{Ashtekar:2011ck,Heinzle:2009du,Montani:2007vu}). However,
what has been achieved so far is  not satisfactory for us as we
wish to work out a classical formalism that would be suitable for
the quantization method that we developed recently (see, e.g.,
\cite{Dzierzak:2009ip,Malkiewicz:2009qv,Malkiewicz:2010py,Mielczarek:2011mx,
Malkiewicz:2011sr,Mielczarek:2012qs}).

In Sec. II we identify the Hamiltonian of our system, which is the
dynamical constraint, and we show that the Lagrangian and
Hamiltonian equations are equivalent. In Sec. III we analyze the
structure of the phase space by using a dynamical system method
and identify the subspace of critical points. It turns out to be
of {\it nonhyperbolic} type. The main part of our paper is
exhibited in Sec. IV. We present the construction of the
symplectic 2-form on the physical phase space and define the
generator of the evolution of our cosmological system, i.e. the
physical Hamiltonian. We make a test of our method on the specific
region of the physical phase space, chosen in such a way that all
calculations can be done comparatively easily.  We conclude in
Sec. VII. Appendix A presents a useful formula to be used in Sec.
V. Appendix B  presents an analysis of dynamics in terms of the
McGehee variables, and confirms that the nonhyperbolicity  is a
generic feature of the considered dynamics.

\section{Equations of motion}

\subsection{Derivation of equations of motion}

In what follows we use the terminology and notation of Refs.
\cite{BKL22,BKL33,Belinski:2009wj,BKL2}.  Our simplified
derivation of dynamics is based on the full derivation presented
in \cite{BKL2}.

The general form of a line element of the nondiagonal Bianchi IX
model, in the synchronous reference system, reads
\begin{equation}\label{dd1}
ds^2 = dt^2 - \gamma_{ab}(t)e^a_\alpha e^b_\beta dx^\alpha
dx^\beta ,
\end{equation}
where Latin indices $a,b,\ldots$ run from $1$ to $3$ and label
frame vectors, and Greek indices $\alpha,\beta,\ldots$ take values
$1,2,3$ and concern space coordinates, and where $\gamma_{ab}$ is
a spatial metric.

The spacetime coordinates and the gauge chosen in the line element
(1) allow the synchronization of clocks at different points of
space. The time coordinate is the proper time at each point of
space. This form of the metric  has been used in Refs.
\cite{BKL22,BKL33,Belinski:2009wj,BKL2,Bogo}, and we apply it
first for consistency. Later we use a different gauge that is more
convenient for our purposes. The choice of the gauge (specifying
time) which leads to the simplification of Einstein's equations is
a common procedure in the analysis of classical dynamics of
gravitational systems. In the Hamiltonian formulation of the
dynamics, different choices of time (which may differ from the
metric time)  lead to different physical Hamiltonians (see Sec.
IV). The choice of an evolution parameter has serious consequences
when one tries to {\it quantize} the dynamics (see, e.g.,
\cite{Malkiewicz:2011sr} for more details).

The homogeneity of the Bianchi IX model means that the three
independent differential 1-forms $e^a_\alpha dx^\alpha$ are
invariant under the transformations of the isometry group of the
Bianchi IX model, i.e. $e^a_\alpha (x) dx^\alpha = e^a_\alpha
(x^\prime) {dx^\prime}^\alpha$, so $e^a_\alpha$ have the same
functional form in old and new coordinate systems. Each
$e^a_\alpha dx^\alpha$ cannot be presented in the form of a total
differential of a function of coordinates. The spatial metric
$\gamma_{ab}(t)$ is a $3\times3$ matrix with all nonvanishing
elements to be determined. The Einstein equations, for the Bianchi
IX model with the metric (\ref{dd1}), become simpler if we
redefine the cosmological time variable $t$ as follows:
\begin{equation}\label{dd2}
    dt = \sqrt{\gamma}\; d\tau ,
\end{equation}
where $\gamma$ denotes the determinant of $\gamma_{ab}$.

It was shown in \cite{BKL22,BKL33,BKL2} that near the cosmological
singularity the {\it general} form of the metric $\gamma_{ab}$
should be considered. Consequently, one cannot globally
diagonalize the metric, i.e.  for all values of time\footnote{In
the case of the Bianchi IX model without the motion of matter,  it
was proved in \cite{Bogo} that the vanishing of the Ricci tensor
components $R^0_a$ (where the index $0$ denotes time coordinate)
enables the diagonalization of the metric $\gamma_{ab}$.}. After
making use of the Bianchi identities, freedom in the rotation of
the metric $\gamma_{ab}$ and frame vectors $e^a_\alpha$, one
arrives at the well-defined but complicated system of equations
specifying the dynamics of the {\it nondiagonal} Bianchi IX model
[see Eqs. (2.14)-(2.20) in \cite{BKL2}].  The assumption that the
{\it anisotropy} of space may grow without bound  when approaching
the singularity [see Eq. (2.21) in \cite{BKL2}] is well justified
[see considerations following Eq. (2.32) in \cite{BKL2}]. It
enables further simplification of the dynamics. Finally, the {\it
asymptotic} form (very close to the cosmological singularity) of
the dynamical equations of the nondiagonal Bianchi IX model is
found to be [see  Eqs. (2.25)-(2.28) of \cite{BKL2}]

\begin{equation}\label{L1} \frac{\partial^2 \ln a }{\partial
\tau^2} = \frac{b}{a}- a^2,~~~~\frac{\partial^2 \ln b }{\partial
\tau^2} = a^2 - \frac{b}{a} + \frac{c}{b},~~~~\frac{\partial^2 \ln
c }{\partial \tau^2} = a^2 - \frac{c}{b},
\end{equation}
where $a,b,c$ are functions of time $\tau$ only.  The solutions to
(\ref{L1}) must satisfy the condition
\begin{equation}\label{L2}
\frac{\partial\ln a}{\partial\tau}\frac{\partial\ln
b}{\partial\tau} + \frac{\partial\ln
a}{\partial\tau}\frac{\partial\ln c}{\partial\tau} +
\frac{\partial\ln b}{\partial\tau}\frac{\partial\ln
c}{\partial\tau} = a^2 + \frac{b}{a} + \frac{c}{b}.
\end{equation}

It is worth mentioning \cite{VB} that one cannot obtain from the
system (\ref{L1})-(\ref{L2}) the dynamics corresponding to the
diagonal case, which was possible in the more general case  [prior
to making the assumption defined by  (2.21) in \cite{BKL2}].

It is easy to verify that (\ref{L1}) can be obtained from the
Lagrangian equations of motion:
\begin{equation}\label{L3}
    \frac{d}{d\tau}\Big(\frac{\partial L}{\partial \dot{x}_I}
    \Big) = \frac{\partial L}{\partial x_I},~~~~I=1,2,3,
\end{equation}
where $x_1 := \ln a,~x_2 := \ln b,~x_3 := \ln c,$ and $\dot{x}_I
:= d x_I/d\tau$, and where  the Lagrangian $L$ has the form
\begin{equation}\label{L4}
L := \dot{x}_1 \dot{x}_2 + \dot{x}_1 \dot{x}_3 + \dot{x}_2
\dot{x}_3 +\exp(2 x_1) + \exp (x_2 -x_1)+ \exp (x_3 -x_2) .
\end{equation}
In what follows we rewrite the dynamics in terms of the
Hamiltonian system.

\subsection{Hamiltonian}

The momenta,  $p_I := \partial L/\partial \dot{x}_I$, are easily
found to be
\begin{equation}\label{H1}
p_1 = \dot{x}_2 + \dot{x}_3,~~~p_2 = \dot{x}_1 + \dot{x}_3,~~~p_3
= \dot{x}_1 + \dot{x}_2 .
\end{equation}
The Hamiltonian of the system has the form
\begin{equation}\label{H2}
H := p_I \dot{x}_I - L = \frac{1}{2}(p_1 p_2 + p_1 p_3 + p_2 p_3)
- \frac{1}{4} (p_1^2 + p_2^2 + p_3^2) - \exp (2 x_1) - \exp (x_2
-x_1)- \exp (x_3 -x_2),
\end{equation}
which due to the relations (\ref{H1}) and  (\ref{L2}) leads to the
dynamical constraint
\begin{equation}\label{H33}
    H = 0.
\end{equation}
The Hamilton equations
\begin{equation}\label{H4}
 \dot{x_I}= \{x_I, H\} = \partial H/ \partial p_I,~~~\dot{p_I}= \{p_I, H\}=
 - \partial H/ \partial q_I,
\end{equation}
where
\begin{equation}\label{pois1}
\{\cdot,\cdot\}:=  \sum_{I=1}^3\Big(\frac{\partial \cdot}
    {\partial x_I} \frac{\partial \cdot}{\partial p_I} -
     \frac{\partial \cdot}{\partial p_I} \frac{\partial \cdot}{\partial
     x_I}\Big),
\end{equation}
have the following explicit form:
\begin{eqnarray}
  \dot{x}_1 &=& \frac{1}{2} (-p_1 + p_2 +p_3), \label{x1}\\
 \dot{x}_2 &=& \frac{1}{2} (p_1 - p_2 + p_3),\label{x2}\\
\dot{x}_3 &=& \frac{1}{2} (p_1 + p_2 - p_3),\label{x3}\\
\dot{p}_1 &=& 2\exp(2 x_1) - \exp (x_2 -x_1), \label{p1}\\
\dot{p}_2 &=& \exp (x_2
-x_1)- \exp (x_3 -x_2), \label{p2}\\
\dot{p}_3 &=& \exp (x_3 -x_2).\label{p3}
\end{eqnarray}
Taking derivatives of (\ref{x1})-(\ref{x3}) and making use of
(\ref{p1})-(\ref{p3}) leads directly to Eq. (\ref{L1}). Since the
constraint (\ref{H33}) is a direct consequence of the constraint
(\ref{L2}), the Lagrangian and Hamiltonian formulations are
completely equivalent.

The system (\ref{x1})-(\ref{p3}) presents a set of {\it nonlinear}
coupled differential equations. The space of the solution of the
above dynamical system is  defined in  $\dR^6$. This space is
bounded by the constraint equation (\ref{H33}). Solving
(\ref{H33})  with respect to $x_3$ gives
\begin{equation}\label{e3}
x_3= x_2+\log \left[-e^{2
x_1}-e^{-x_1+x_2}-\frac{p_1^2}{4}+\frac{p_1
p_2}{2}-\frac{p_2^2}{4}+\frac{p_1 p_3}{2}+\frac{p_2
p_3}{2}-\frac{p_3^2}{4}\right].
\end{equation}
Substituting (\ref{e3}) into (\ref{x1})-(\ref{p3}) we get \bea
\dot{x}_1 &=&\frac{1}{2} (-p_1+p_2+p_3),\label{dx1}\\
\dot{x}_2 &=& \frac{1}{2} (p_1-p_2+p_3),\\
\dot{p}_1 &=&2 e^{2 x_1}-e^{-x_1+x_2},\\
\dot{p}_2 &=&e^{2 x_1}+2 e^{-x_1+x_2}+\frac{p_1^2}{4}-\frac{p_1
p_2}{2}+
\frac{p_2^2}{4}-\frac{p_1 p_3}{2}-\frac{p_2 p_3}{2}+\frac{p_3^2}{4},\\
\dot{p}_3 &=&-e^{2 x_1}-e^{-x_1+x_2}-\frac{p_1^2}{4}+\frac{p_1
p_2}{2}-\frac{p_2^2}{4}+\frac{p_1 p_3}{2}+\frac{p_2
p_3}{2}-\frac{p_3^2}{4}\label{dp3}. \eea

As far as we are aware, an analytical solution to this system  is
unknown. In what follows we try to understand   {\it local}
properties of the phase space of the system by using the method of
nonlinear dynamical systems \cite{Arnold,Perko,Wiggins,
Czuchry:2009hz,Czuchry:2010vx}.

\section{Dynamical systems analysis of phase space}

The local geometry of the phase space is characterized by the
nature and position of its {\it critical} points\footnote{ In what
follows we use the terms critical or fixed points.}. These points
are locations where the derivatives of all the dynamical variables
vanish. These  are the points where phase trajectories may start,
end, intersect, etc. The trajectories can also begin or end at
infinities. Then, after a suitable coordinate transformation that
maps  an unbounded phase space into a compact region $S$,  these
originally critical points at infinity may become better defined
for further analysis. However, such a procedure usually leads to
the situation where the dynamical system is mapped into the
interior of $S$ and the critical points at infinity are mapped
into the boundary  of $S$, which may consist of faces intersecting
at corners of various dimensions. Consequently, an identification
of separatrices connecting the critical points may become
challenging. Apart from these, the compactification mapping is
usually a noncanonical transformation that may lead to a
Hamiltonian system that is different from the original one. The
compactification is usually a very subtle procedure
\cite{Bogo,Czuchry:2009hz,Czuchry:2010vx}. The set of finite and
infinite critical points and their characteristics, given by the
properties of the Jacobian matrix of the linearized equations at
those points, may provide a {\it qualitative} description of a
given dynamical system.

The above situation is specific to the case where a fixed  point
is of the {\it hyperbolic} type. In the case of the {\it
nonhyperbolic} fixed point, a {\it linearized} vector field at the
fixed point cannot be used to characterize {\it completely} the
local properties of the phase space.

\subsection{Critical points of the ``unconstrained'' system}

Inserting $\dot{x}_1 = 0 = \dot{x}_2 =\dot{x_3}=\dot{p}_1=
\dot{p}_2 = \dot{p}_3$ into the left-hand sides of
(\ref{x1})-(\ref{p3}) leads to the following set of equations:
\begin{eqnarray}
 0 &=& \frac{1}{2} (-p_1 + p_2 +p_3), \label{xx1}\\
0 &=& \frac{1}{2} (p_1 - p_2 + p_3),\label{xx2}\\
0 &=& \frac{1}{2} (p_1 + p_2 - p_3),\label{xx3}\\
0 &=& 2\exp(2 x_1) - \exp (x_2 -x_1), \label{pp1}\\
0 &=& \exp (x_2
-x_1)- \exp (x_3 -x_2), \label{pp2}\\
0 &=& \exp (x_3 -x_2),\label{pp3}
\end{eqnarray}
plus the Hamiltonian constraint equation
\begin{equation}\label{H3}
 \frac{1}{2}(p_1 p_2 + p_1 p_3 + p_2 p_3) - \frac{1}{4} (p_1^2 +
p_2^2 + p_3^2) - \exp (2 x_1) - \exp (x_2 -x_1)- \exp (x_3 -x_2) =
0.
\end{equation}
Equations (\ref{xx1})-(\ref{xx3}) yield \be
p_1=0=p_2=p_3,\label{ppp} \ee whereas Eqs. (\ref{pp1})-(\ref{pp3})
lead to the conditions \bea
0 &=& \exp(2 x_1) , \label{ppx1}\\
0 &=& \exp (x_2
-x_1), \label{ppx2}\\
0 &=& \exp (x_3 -x_2),\label{ppx3} \eea Conditions
(\ref{ppx1})-(\ref{ppx3}) are fulfilled for \bea
x_1&\rightarrow&-\infty,\\
x_2-x_1&\rightarrow&-\infty,\\
x_3-x_2&\rightarrow&-\infty. \eea

Thus the set  of critical points $S_B$ is given by \bea
\label{critS}S_B: &=& \{(x_1,x_2,x_3,p_1,p_2,p_3)\in \bar{\dR}^6
~|~ (x_1 \rightarrow
    -\infty,~ x_2-x_1 \rightarrow -\infty,~ x_3-x_2 \rightarrow -\infty)\nonumber
    \\&& \wedge (p_1 = 0 = p_2 = p_3  \label{critS0}\}, \eea
where $\bar{\dR}:= \dR \cup  \{-\infty, +\infty\}$. It is not easy
to give a more specific definition of $S_B$, situated at
``infinity'', with our choice of the phase space variables. The
``direction'' and the way it is approached should be taken into
account. Moreover, the equations of motion should also be
respected. One may speculate that the infinities in (\ref{critS0})
should be approached in such a way that $x_3 \ll x_2 \ll x_1$.
Much better way of dealing with the points at infinities would
consist in an appropriate mapping of our phase space onto a
compact region of $\dR^6$ by using, for instance, the
Poincar\'e-type mapping \cite{Czuchry:2009hz,Czuchry:2010vx}.
Since in the present work we do not intend to analyze the set
$S_B$ in terms of possible separatrices, we postpone making use of
the compactification maps to our next paper.

The stability properties are determined by the eigenvalues of the
Jacobian of the system (\ref{x1})-(\ref{p3}). More precisely, one
has to linearize Eqs. (\ref{x1})-(\ref{p3}) at each point.
Inserting $\vec{x}=\vec{x}_0+\delta\vec{ x}$, where
$\vec{x}=(x_1,x_2,x_3,p_1,p_2,p_3)$, and keeping terms up to first
order in $\delta\vec{x} $ leads to an evolution equation of the
form $\delta\dot{\vec{ x}}=J\delta\vec{x}$. Eigenvalues of $J$
describe stability properties at the given point.

The Jacobian $J$ of the system (\ref{x1})-(\ref{p3}), evaluated at
any point of $S_B$, reads:
\begin{equation}
\nonumber J=\left(\begin{array}{cccccc}
    0 & 0  & 0  & -1/2 &  1/2 &  1/2 \\
    0 & 0  & 0  &  1/2 & -1/2 &  1/2 \\
        0 & 0  & 0  &  1/2 & 1/2  & -1/2 \\
    0 & 0 & 0 & 0    & 0    & 0    \\
       0 & 0  & 0 & 0    & 0    & 0    \\
       0 & 0 & 0  & 0    & 0    & 0
\end{array} \right)
\end{equation}
The characteristic polynomial associated with Jacobian $J$ is: \be
P(\lambda)=\lambda^6, \ee so the  eigenvalues are the following:
\be \left(0,0,0,0,0,0\right). \ee

Since the real parts of all eigenvalues of the Jacobian are equal
to zero, the set (\ref{critS}) consists of {\it nonhyperbolic}
fixed points.

A critical point is called a hyperbolic fixed point if {\it all}
the eigenvalues of the Jacobian matrix of the linearized equations
at this point have nonzero real parts. Otherwise, it is called a
nonhyperbolic fixed point \cite{Perko,Wiggins}.

\subsection{Critical points of the ``constrained'' system}

Inserting $\dot{x}_1 = 0 = \dot{x}_2 =\dot{p}_1= \dot{p}_2 =
\dot{p}_3$ into the left-hand side  of (\ref{dx1})-(\ref{dp3})
leads to the following set of equations: \bea
0 &=&\frac{1}{2} (-p_1+p_2+p_3),\label{dxx1}\\
0 &=& \frac{1}{2} (p_1-p_2+p_3),\label{dxx2}\\
0 &=&2 e^{2 x_1}-e^{-x_1+x_2},\label{dpp1}\\
0 &=&e^{2 x_1}+2 e^{-x_1+x_2}+\frac{p_1^2}{4}-\frac{p_1 p_2}{2}+
\frac{p_2^2}{4}-\frac{p_1 p_3}{2}-\frac{p_2 p_3}{2}+\frac{p_3^2}{4},\label{dpp2}\\
0 &=&-e^{2 x_1}-e^{-x_1+x_2}-\frac{p_1^2}{4}+\frac{p_1
p_2}{2}-\frac{p_2^2}{4}+\frac{p_1 p_3}{2}+\frac{p_2
p_3}{2}-\frac{p_3^2}{4}\label{dpp3}. \eea

Equations (\ref{dxx1})-(\ref{dxx2}) yield
\begin{equation}\label{dxxx}
    p_1 = p_2,~~~~p_3 =0.
\end{equation}

Equations (\ref{dpp1})-(\ref{dpp3}) and the above solution
(\ref{dxxx}) lead to the following set of equations: \bea
0 &=&2 e^{2 x_1}-e^{-x_1+x_2},\label{ddpp1}\\
0 &=&e^{2 x_1}+2 e^{-x_1+x_2},\label{ddpp2}\\
0 &=&-e^{2 x_1}-e^{-x_1+x_2}\label{ddpp3}.\eea
Solutions of the
above set are the following: \bea
x_1&\rightarrow&-\infty,\\
x_2-x_1&\rightarrow&-\infty.\\
\eea Taking into account the Hamiltonian constraint (\ref{H2})
leads to the additional condition \be
x_3-x_2\rightarrow-\infty.\ee Requiring that the time derivative
of $x_3$, Eq (\ref{e3}), determined from the Hamiltonian
constraint (\ref{H2}) satisfies (\ref{x3}) leads to Eq
(\ref{xx3}), which together with (\ref{dxx1}) and (\ref{dxx2})
finally gives $p_1 = 0 = p_2 = p_3$ .

In this way we have shown that in both cases  of constrained and
unconstrained dynamics the set of critical points $S$ is defined
by Eq. (\ref{critS}).

The Jacobian $J$ of the system (\ref{dx1})-(\ref{dp3}) evaluated
at the set of critical points $S$ reads
\begin{equation}
\nonumber J=\left(\begin{array}{cccccc}
    0 & 0    & -1/2 &  1/2 &  1/2 \\
    0 & 0    &  1/2 & -1/2 &  1/2 \\
        0 & 0 & 0     & 0    & 0    \\
       0 & 0  & 0    & 0    & 0    \\
       0 & 0  & 0    & 0    & 0
\end{array} \right)
\end{equation}
The characteristic polynomial associated with Jacobian $J$ is
equal to \be P(\lambda)=\lambda^5, \ee so the  eigenvalues are the
following: \be \left(0,0,0,0,0\right) \ee which is consistent with
the result obtained for the case of the unconstrained dynamics.

In both cases of constrained and unconstrained dynamics we are
dealing with the  nonhyperbolic type of critical points. Thus,
getting insight into the structure of the space of orbits near
such points requires an examination of the {\it exact} form of the
vector field defining the phase space of our dynamical system. The
information obtained from linearization is inconclusive.

The results we have obtained so far have the following properties:
\begin{enumerate}
    \item The phase space is six dimensional.
    \item The critical points we are dealing with are of
    {\it nonhyperbolic} type.
    \item The set of critical points $S_B$, defined by (\ref{critS}), is not a set of
    isolated points, but a three-dimensional {\it continuous} subspace of $\bar{\dR}^6$.
    \item The critical subspace $S_B$ belongs to an {\it asymptotic} region of phase
    space with infinite values of its variables.
\end{enumerate}

Lower-dimensional phase spaces can be easily analyzed by making
mapping to known solved cases available in textbooks. Our
six-dimensional phase space requires sophisticated  tools. The
second  property is challenging. The nonhyperbolicity means that
one cannot avoid direct examination of the original {\it
nonlinear} set of equations defining the dynamics. The
corresponding linearized set of equations is unable to reveal the
nature of dynamics in the neighborhoods of  fixed points. The
vector field may, e.g.  bifurcate there \cite{Perko,Wiggins}).
{\it Nonisolated} critical points are atypical ones. The standard
methods presented in textbooks \cite{Arnold,Perko,Wiggins} cannot
cope easily with such a case.  Making perturbations around
critical points at infinities is not a well-defined procedure.
However, this problem may be avoided by mapping unbounded
subspaces onto {\it finite} ones, so the fourth property should
not lead to serious problems.

\section{Hamiltonian structure of reduced system}

In what follows we propose a new theoretical framework.  We turn
our system with the Hamiltonian constraint into a new dynamical
system.  In this new system the Hamiltonian is no longer a
constraint, but a generator of an evolution. We call it a {\it
true} (physical) Hamiltonian.

Solutions to Eqs.  (\ref{x1})-(\ref{p3}) define the {\it
kinematical} phase space $\mathcal{F}_k$. The {\it physical} phase
space $\mathcal{F}_p$ is defined by those solutions to
(\ref{x1})-(\ref{p3}) which additionally satisfy the constraint
defined by (\ref{H33}). The physical Dirac {\it observable}
$\mathcal{O}$ is defined to be a function on the physical phase
space that {\it weakly} Poisson commutes with first-class
constraints of the dynamical system \cite{PAM,HT,Dzierzak:2009ip}.
In our case there is only one constraint (\ref{H33}) so the
equation defining the physical Dirac observables reads
\begin{equation}\label{dir}
    \{\mathcal{O},H\}\approx 0,~~~~~\mbox{where}~~~\{\cdot,\cdot\}:=  \sum_{k=1}^3
    \Big(\frac{\partial \cdot}
    {\partial x_k} \frac{\partial \cdot}{\partial p_k} -
     \frac{\partial \cdot}{\partial p_k} \frac{\partial \cdot}{\partial
     x_k}\Big),
\end{equation}
and where the sign $\,\approx \,$ means that (\ref{dir}) has to be
treated as a weak equation in Dirac's sense \cite{PAM}. Functions
which satisfy (\ref{dir}) {\it strongly}  define the {\it
kinematical} observables. The {\it physical} observable can be
obtained from the solutions to Eq. (\ref{dir}) satisfying the
constraint (\ref{H3}).  The physical Dirac observables are in fact
{\it integrals} of motion. {\it Constants}  of motion are called
{\it partial}  physical Dirac observables. They become the
integrals of motion at fixed values of an evolution parameter
(time) of the system.

Let us  rewrite the classical dynamics in terms of the true
Hamiltonian.  To begin with, we consider  the following {\it
factorization}, consistent with the Darboux theorem \cite{AM,NMJ}:
\begin{equation}\label{factor}
\omega := \sum_{k=1}^3 \Big(dx_k \wedge dp_k \Big) =  \sum_{\alpha
=1}^2 \Big(d\tilde{q}_\alpha \wedge d\tilde{\pi}_\alpha\Big) +
d\tilde{t}\wedge dH,
\end{equation}
where $(p_k,x_k) \in \mathcal{F}_k$ and
$(\tilde{q}_\alpha,\tilde{\pi}_\alpha)$ are some new canonical
variables; $H\neq 0$ is an extension of $H$ to the neighborhood of
the constraint surface defined by $H=0$. It is clear that the
transformation $(x_1,x_2,x_3,p_1,p_2,p_3) \rightarrow
(\tilde{q}_1,\tilde{q}_2,\tilde{t},\tilde{\pi}_1,\tilde{\pi}_2,H)$
underlying (\ref{factor}) is canonical.

We  wish to map the symplectic structure (\ref{factor}) of the
kinematical level, defined in $\mathcal{F}_k$,  into a Hamiltonian
structure in $\mathcal{F}_p$. Let us consider
\begin{equation}\label{facHH}
\Omega := \omega_{|_{H=0}} = \sum_{k=1}^3 \Big(dx_k \wedge dp_k
\Big)_{|_{H=0}}.
\end{equation}

Suppose (\ref{facHH}) can be rearranged  as follows:
\begin{equation}\label{facH}
\Omega  = \sum_{\alpha =1}^2 \Big(dq_\alpha \wedge
d\pi_\alpha\Big) + dT\wedge dH_T,
\end{equation}
where $q_\alpha,\pi_\alpha$ and $T $ are new canonical variables,
and where $H_T = H_T(q_\alpha,\pi_\alpha, T)$. We emphasize that
the existence of the expression (\ref{facH}) cannot be guaranteed
in advance. However, let us assume that (\ref{facH}) can be
constructed. Later, we examine the existence problem of such a
2-form in terms of the Pfaff equation \cite{Pfaff}.  Due to the
factorization (\ref{facH}) and considerations on time-dependent
vector fields and differential forms in
\cite{AM,NMJ,Malkiewicz:2011sr}, we obtain
\begin{equation}\label{D1}
\frac{d}{dT}q_\alpha := \{q_\alpha,H_T\}_{q, \pi} = \frac{\partial
H_T}{\partial \pi_\alpha}
\end{equation}
and
\begin{equation}\label{D2}
\frac{d}{dT}\pi_\alpha := \{\pi_\alpha,H_T\}_{q, \pi} =
-\frac{\partial H_T}{\partial q_\alpha} ,
\end{equation}
where
\begin{equation}\label{D3}
\{\cdot,\cdot\}_{q, \pi}:=  \sum_{\alpha =1}^2 \Big(\frac{\partial
\cdot}
    {\partial q_\alpha} \frac{\partial \cdot}{\partial \pi_\alpha} -
     \frac{\partial \cdot}{\partial \pi_\alpha} \frac{\partial \cdot}{\partial
     q_\alpha}\Big) .
\end{equation}
Therefore, the existence  of (\ref{facH}) implies   the existence
of the Hamiltonian structure of the reduced system.  The dynamics
is generated by the  true Hamiltonian $H_T$ and is parametrized by
an evolution parameter $T$.

In this new setting the system has no dynamical constraints. One
may examine it using the dynamical system methods
\cite{Perko,Wiggins}. The phase space is now only {\it four}
dimensional, which simplifies the analysis of the original
dynamics, (\ref{x1})-(\ref{p3}), defined in {\it six}-dimensional
phase space with the dynamical constraint. The factorization
procedure described above makes sense {\it outside} the subspace
of the critical points $S_B$ defined by (\ref{critS}) for the
reason given later when we try to solve  the dynamics
analytically.

\section{Absolute dynamics}

Suppose we are able to solve analytically EQs.
(\ref{x1})-(\ref{e3}) {\it outside} of the critical space $S_B$.
As a result, we obtain the phase space variables as functions of
constants of motion and time (\ref{dd2}). Now, the time has a
direct link with the metric (\ref{dd1}). We call it an {\it
absolute} time.

In the next step, we insert the phase space variables into
(\ref{facHH}) and try to get a  2-form as  a prerequisite for the
construction of a Hamiltonian system.   As soon as we identify
this structure, we may try to make an {\it extension} of it to the
space $S_B$. This way, we may analyze the structure of $S_B$.

In what follows, we realize, to some extent,  this idea. We solve
the dynamics in some approximation and in some special subspace of
the physical phase space, and find the corresponding Hamiltonian
structure.

\subsection{Analytical solution of the dynamics}

Let us find an {\it approximate} analytic solution to the
dynamical equations (\ref{x1})-(\ref{p3}) by using the method of
successive approximations (see, e.g. \cite{Perko,KM}).

The system (\ref{x1})-(\ref{p3}) can be rewritten in the form
\begin{equation}\label{as1}
\frac{d}{d\tau} \left(\begin{array}{c} x_1 \\ x_2 \\ x_3\\p_1 \\ p_2 \\
p_3\end{array} \right) =
\left(\begin{array}{c} (-p_1+p_2+p_3)/2  \\
 (p_1-p_2+p_3)/2 \\
 (p_1 + p_2-p_3)/2\\
2 e^{2 x_1}-e^{x_2 -x_1} \\
e^{x_2 - x_1} - e^{x_3 - x_2}\\
e^{x_3 - x_2}
\end{array} \right) ,
\end{equation}
where the solutions to (\ref{as1}) must satisfy the constraint
\begin{equation}\label{as2}
x_3= x_2+\log \left[-e^{2 x_1}-e^{-x_1+x_2}-\frac{1}{4}(p_1^2 +
p_2^2 + p_3^2)+\frac{1}{2}(p_1 p_2 + p_1 p_3 + p_2 p_3)\right].
\end{equation}
One may verify that outside of the set $S_B$  of the critical
points (\ref{critS}), the initial value problem for Eq.
(\ref{as1}) has a {\it unique} solution.  This is so because
outside of $S_B$ the right-hand side of (\ref{as1}) defines a
mapping that is continuous and bounded and that satisfies the
Lipschitz condition. The initial value condition
\begin{equation}\label{as3}
\big(x_1(\tau_0),x_2(\tau_0),x_3(\tau_0),p_1(\tau_0),p_2(\tau_0),p_3(\tau_0)\big)
= \big( x_{10}, x_{20},x_{30},p_{10},p_{20},p_{30} \big)=: y_0
\end{equation}
must satisfy the constraint (\ref{as2}).

The Cauchy problem defined by (\ref{as1}) and (\ref{as3}) can be
written in a compact way as follows:
\begin{equation}\label{as4}
\frac{d}{d\tau}y(\tau) = f(y(\tau)),~~~~y(\tau_0) = y_0,~~~~y \in
\dR^6,~~~~f(y) \in \dR^6,
\end{equation}
which is equivalent to the equation
\begin{equation}\label{as5}
y(\tau) = y_0 +\int_{\tau_0}^\tau dt \;f(y(t)).
\end{equation}
Equation (\ref{as5}) can be solved by an iteration.  The iterated
solutions satisfy:
\begin{equation}\label{as6}
y^{(n)}(\tau) = y_0 +\int_{\tau_0}^\tau dt \;f(y^{(n-1)}(t)),
\end{equation}
where $y_0 \notin S_B$ and is defined by (\ref{as3}).

 The more iterations we make, the better approximation to the
exact solution we obtain. In what follows we solve (\ref{as6}) in
the first iteration approximation.

The simplest ``iteration'' reads $(x_1,x_2,x_3,p_1,p_2,p_3)=
(x_{10},x_{20},x_{30},p_{10},p_{20},p_{30})$, where  due to
(\ref{as2}) we must have
\begin{equation}\label{x30}
x_{30}= x_{20}+\log \left[-e^{2 x_{10}}- e^{-x_{10}+x_{20}}
-\frac{1}{4}(p^2_{10}+ p^2_{20} +p^2_{30})
+\frac{1}{2}(p_{10}p_{20} + p_{10}p_{30} + p_{20}p_{30}) \right].
\end{equation}

The first iteration of  (\ref{as6}) gives
\begin{equation}\label{aas1}
 \left(\begin{array}{c} x_1 \\ x_2 \\ x_3\\p_1 \\ p_2 \\
p_3\end{array} \right) =
\left(\begin{array}{c}x_{10}+\frac{1}{2} (-p_{10} + p_{20} + p_{30}) (\tau - \tau_0)  \\
x_{20}+\frac{1}{2}  (p_{10} - p_{20} + p_{30}) (\tau - \tau_0) \\
x_{30}+\frac{1}{2}  (p_{10} + p_{20} - p_{30}) (\tau - \tau_0) \\
 p_{10} + (2 e^{2 x_{10}} - e^{x_{20} -x_{10}}) (\tau - \tau_0) \\
p_{20} + \big(e^{x_{20}- x_{10}} - e^{x_{30}- x_{20}} \big)(\tau - \tau_0)\\
p_{30} + e^{x_{30}- x_{20}}(\tau - \tau_0)
\end{array} \right) ,
\end{equation}
and the solution (\ref{aas1}) must respect the constraint
(\ref{x30}).

\subsection{Determination of the true Hamiltonian}

Since the expression for the 2-form (\ref{facH}) consists of
differentials, we should first find the relationship
\begin{equation}\label{relAB}
(dx_1, dx_2, dx_3,dp_1, dp_2, dp_3)~~~\longleftrightarrow
~~~(dx_{10},dx_{20},dx_{30},dp_{10},dp_{20},dp_{30}),
\end{equation}
in such a way that the constraint (\ref{as2}) is satisfied. The
successive iterations may lead far away from the constraint
surface. Thus, each level of the iteration procedure should be
preceded by an imposition of the constraint. We have performed
only one step of the iteration. To simplify calculations, we
impose the constraint only on the right-hand side of (78). This
means that we assume that the first iteration  does not destroy
{\it much} the constraint imposed on the solution at $\tau =
\tau_0$. This assumption seems to be plausible outside of the
critical subspace $S_B$ defined by (41), and for an infinitesimal
$|\tau - \tau_0|$. This way, we get some insight into our method,
and obtain an approximate solution to Eq. (\ref{as5}).

To get insight into the above scheme, we consider a special case
defined by the conditions
\begin{equation}\label{sim34}
 p_{30}= p_{10}-p_{20},~~~~q_2 - q_1 \rightarrow -\infty,
\end{equation}
where $q_1 := x_{10}\rightarrow -\infty$ and $q_2 :=
x_{20}\rightarrow -\infty$.

In the region defined by (\ref{sim34}) the 2-form $\Omega$,
determined in Appendix A, reads
\begin{equation}\label{sim4}
    2 \Omega = \Big(- \pi_2^2 \;d\pi_1 +(\pi_1^2 - 2\pi_1 \pi_2 +
    2\pi_2^2)\;d\pi_2\Big)\wedge dT ,
\end{equation}
where $\pi_1 := p_{10}$ and $\pi_2 := p_{20}$. Our intention is to
rewrite (\ref{sim4}) in the form $\Omega = dH_T \wedge dT$, which
means that we wish to find  $H_T = H_T(\pi_1, \pi_2)$ such that
\begin{equation}\label{sim5}
dH_T = \frac{\partial H_T}{\partial \pi_1}d\pi_1 + \frac{\partial
H_T}{\partial \pi_2}d\pi_2 = - \pi_2^2 \;d\pi_1 +(\pi_1^2 - 2\pi_1
\pi_2 + 2\pi_2^2)\;d\pi_2 .
\end{equation}
Equation (\ref{sim5}) has the form of the Pfaff equation
\cite{Pfaff}. Finding $H_T$ that satisfies (\ref{sim5}) means
finding a {\it complete} integral of this equation. It is commonly
known that the necessary and sufficient condition for the
existence of such an integral reads
\begin{equation}\label{sim6}
\frac{\partial^2 H_T}{\partial \pi_1 \partial \pi_2} =
\frac{\partial^2 H_T}{\partial \pi_2 \partial \pi_1}.
\end{equation}
One can easily verify that the coefficients of $d\pi_1$ and
$d\pi_2$  in the right-hand side of (\ref{sim5})  do not satisfy
the condition (\ref{sim6}). This means that the problem of the
integrability of (\ref{sim5}) may have a solution, but in the case
where $\pi_1 = \pi_1(\pi_2)$ or $\pi_2 = \pi_2(\pi_1)$. In fact
there exists an infinite number of solutions with lost
independence of $\pi_1$ and $\pi_2$. Since we wish to find  $H_T =
H_T(\pi_1,\pi_2)$, we rely on the method that uses an idea of an
integrating factor $\mu = \mu(\pi_1, \pi_2)$. Making use of this
leads to a new equation for $H_T$:
\begin{equation}\label{sim7}
dH_T = \mu(\pi_1, \pi_2) \big(- \pi_2^2 \;d\pi_1 +(\pi_1^2 -
2\pi_1 \pi_2 +
    2\pi_2^2)\;d\pi_2\big).
\end{equation}
Now, the condition (\ref{sim6}) leads to the following equation
for the factor $\mu$:
\begin{equation}\label{sim8}
(\pi_1^2 - 2 \pi_1 \pi_2 + 2 \pi_2^2)\frac{\partial \mu}{\partial
\pi_1} + \pi_2^2 \frac{\partial \mu}{\partial \pi_2} + 2 \pi_1 \mu
= 0.
\end{equation}
Solving (\ref{sim8}) does not lead to a unique expression for
$H_T$. This is because  the new expression $\tilde{\mu}:= f(H_T)
\mu$, where $f$ is any function, is an integrating factor of
(\ref{sim7}) as well. Fortunately, this is not an obstacle since
$dH_T$ occurs in the context of $dH_T \wedge dT$ in the expression
(\ref{sim4}). The ambiguity can be easily absorbed by the
redefinition of time due to the identity $f(H_T) dH_T \wedge dT =
dH_T \wedge f(H_T)dT =: dH_T \wedge d\tilde{T}$.

An example of a solution to (\ref{sim8}), found by the method of
characteristics, is given by:
\begin{equation}\label{sim9}
\mu(\pi_1,\pi_2)=
\frac{1}{(\pi_1-2\pi_2)^2}F\big(\log\frac{2(\pi_2^2-\pi_1\pi_2)}{3(\pi_1-2\pi_2)}\big),
\end{equation}
where $F(\cdot)$ is an arbitrary $C^1$ function.

It is commonly  known that the general solution to a partial
differential equation involves arbitrary functions. They can be
fixed after an imposition of suitable Cauchy-type conditions
\cite{Aris,Poly}. We do not discuss this issue here, as our
intention is only to present the method. One may simply verify
that (\ref{sim9}) satisfies (\ref{sim8}). For subsequent
considerations we make the simplest choice of the function $F$ by
taking $F\equiv 1$. Thus we have
\begin{equation}\label{sim10}
\mu(\pi_1,\pi_2)= \frac{1}{(\pi_1-2\pi_2)^2}.
\end{equation}

Now, let us solve Eq. (\ref{sim7}) with respect to $H_T$.
Insertion of $dH_T = \frac{\partial H_T}{\partial \pi_1} d\pi_1 +
\frac{\partial H_T}{\partial \pi_2} d\pi_2 $ into the left-hand
side of (\ref{sim7}) and comparing the coefficient of $d\pi_1$ and
$d\pi_2$ of both sides of (\ref{sim7})  leads to
\begin{eqnarray}
\partial H_T(\pi_1,\pi_2)/\partial\pi_1 &=& -\pi_2^2\; \mu(\pi_1,\pi_2), \label{sim11}\\
\partial H_T(\pi_1,\pi_2)/\partial\pi_2 &=& (\pi_1^2 -2\pi_1 \pi_2 + 2
\pi_2^2)\; \mu(\pi_1,\pi_2). \label{sim12}
\end{eqnarray}
The solution to the system (\ref{sim11})-(\ref{sim12}) may be
taken as

\begin{equation}\label{sim20}
H_T=\frac{\pi_1\pi_2-\pi^2_2}{\pi_1-2\pi_2}.
\end{equation}
One may simply verify that (\ref{sim20}) satisfies  (\ref{sim11}),
 (\ref{sim12}) and (\ref{sim6}).

\subsection{Dynamics in reduced phase space}

In the region defined by (\ref{sim34}), the dynamics is defined by
Eqs. (\ref{D1}) and (\ref{D2}),  so it reads
\begin{eqnarray}
dq_1/dT &=& \partial H_T(\pi_1,\pi_2)/\partial\pi_1 = -\frac{\pi_2^2}{(\pi_1-2\pi_2)^2}, \label{sim14}\\
dq_2/dT  &=& \partial H_T(\pi_1,\pi_2)/\partial\pi_2 =
\frac{\pi_1^2
-2\pi_1 \pi_2 + 2 \pi_2^2}{(\pi_1-2\pi_2)^2}, \label{sim15}\\
d\pi_1/dT &=& - \partial H_T(\pi_1,\pi_2)/\partial q_1 = 0, \label{sim16}\\
d\pi_2/dT &=& - \partial H_T(\pi_1,\pi_2)/\partial q_2 = 0.
\label{sim17}
\end{eqnarray}
One may think that the point of the region (\ref{sim34}) with
coordinates
\begin{equation}\label{sim18}
  (q_1,q_2,\pi_1,\pi_2) = (-\infty,-\infty,0,0)\in \bar{\dR}^4
\end{equation}
is a fixed point of the vector field (\ref{sim14})-(\ref{sim17}).
However, elementary calculations give
\begin{eqnarray}\label{slim1}
&&\lim_{\pi_1\rightarrow 0}\lim_{\pi_2\rightarrow
0}\;\frac{-\pi_2^2}{(\pi_1-2\pi_2)^2}= 0,\label{slim1}
\\
&& \lim_{\pi_2\rightarrow 0}\lim_{\pi_1\rightarrow
0}\;\frac{-\pi_2^2}{(\pi_1-2\pi_2)^2}= -\frac14, \label{slim2}
\end{eqnarray}
and
\begin{eqnarray}
\lim_{\pi_1\rightarrow 0}\lim_{\pi_2\rightarrow 0}\;\frac{\pi_1^2
-2\pi_1 \pi_2 + 2 \pi_2^2}{(\pi_1-2\pi_2)^2}= 1, \label{slim3}\\
\lim_{\pi_2\rightarrow 0}\lim_{\pi_1\rightarrow 0}\;\frac{\pi_1^2
-2\pi_1 \pi_2 + 2 \pi_2^2}{(\pi_1-2\pi_2)^2}=\frac12 .
\end{eqnarray}
Therefore, the vector field (\ref{sim14})-(\ref{sim17}) has a
discontinuous time derivative at the point (\ref{sim18}). Another
singular point of the vector field (\ref{sim14})-(\ref{sim17}) has
the coordinates:
\begin{equation}\label{sim19}
  (q_1,q_2,\pi_1,\pi_2) = (-\infty,-\infty,2\pi_2,\pi_2)\in \bar{\dR}^4.
\end{equation}
It is clear that the right-hand side of (\ref{sim14}) and
(\ref{sim15}) blow up at this point. We exclude from further
analysis both singular points (\ref{sim18}) and (\ref{sim19}). All
other points are regular, i.e., are not critical.

One can easily solve Eqs. (\ref{sim14})-(\ref{sim17}):
\begin{equation}\label{solfin}
q_1(T) = a_1 T + a_2,~~~q_2(T) = b_1 T + b_2,~~~\pi_1(T)=
c_1,~~~\pi_2(T)= c_2,
\end{equation}
where
\begin{equation}\label{cofin}
a_1 := -\frac{c_2^2}{(c_1-2 c_2)^2},~~~a_2 \in \dR,~~~b_1 :=
\frac{c_1^2 -2c_1 c_2 + 2 c_2^2}{(c_1-2 c_2)^2},~~~b_2 \in \dR,
\end{equation}
and where $0 \neq c_1 \in \dR$ and $0 \neq c_2 \in \dR$,
satisfying $c_1 - 2 c_2 \neq 0.$  Since $c_1\neq 0$ and $c_2\neq
0$, we cannot make an extension of our results to the set $S_B$,
starting from the subspace (\ref{sim34}).

\section{Relative dynamics}

The Hamilton equations (\ref{x1})-(\ref{p3}), due to their origin
(\ref{L1})-(\ref{L2}), concern the dynamics near  the cosmological
singularity,  so they have a physical sense only {\it locally} for
an infinitesimal interval of time. From (\ref{p3}) we obtain
$\dot{p}_3 \neq 0$, for $x_2 < \infty$ and $x_3 < \infty$. Thus,
$p_3$  is {\it locally}  monotonic and can play the role of {\it
relative} time $T$.  In what follows we choose $T := - p_3 $.
Because of  the dynamical constraint (\ref{e3}), the variable
$x_3$ can be expressed in terms of other variables, so it may be
chosen to be $H_T$, according to Eq. (\ref{facH}). Making use of
the above substitution, we get

\begin{equation}\label{sim1}
\Omega = dq_1 \wedge d\pi_1 + dq_2 \wedge d\pi_2 + dT \wedge dH_T,
\end{equation}
where
\begin{equation}\label{sim2}
q_1 := x_1,~~~q_2 := x_2,~~~\pi_1 := p_1,~~~\pi_2 := p_2,~~~T = -
p_3,
\end{equation}
and where
\begin{equation}\label{sim3}
H_T = q_2+\log \left[-e^{2 q_1}-e^{-q_1+q_2}-\frac{1}{4}(\pi_1^2 +
\pi_2^2 +T^2) +\frac{1}{2}(\pi_1 \pi_2 + \pi_1 T + \pi_2
T)\right].
\end{equation}
From (\ref{D1})-(\ref{D3}) we get
\begin{eqnarray}
 \label{ff1} \frac{d q_1}{d T} &=& \frac{\partial H_T}{\partial \pi_1} =  \frac{- \pi_1 +
 \pi_2 +x }{2 F},\\
 \label{ff2} \frac{d q_2}{d T} &=& \frac{\partial H_T}{\partial \pi_2} =  \frac{ \pi_1
  - \pi_2 +x }{2 F},\\
\label{ff3} \frac{d \pi_1}{d T} &=& - \frac{\partial H_T}{\partial q_1} = \frac{2 e^{2 q_1}- e^{-q_1 + q_2}}{F}  ,\\
\label{ff4} \frac{d \pi_2}{d T} &=&- \frac{\partial H_T}{\partial q_2} = - 1 + \frac{e^{-q_1 + q_2}}{F}, \\
\label{ff5} \frac{d x}{d T} &=& 1 ,
\end{eqnarray}
where $x := T$, and where
\begin{equation}\label{FFF}
F(q_1,q_2,\pi_1,\pi_2, T):= -e^{2
q_1}-e^{-q_1+q_2}-\frac{1}{4}(\pi_1^2 + \pi_2^2 +T^2)
+\frac{1}{2}(\pi_1 \pi_2 + \pi_1 T + \pi_2 T)
\end{equation}
The system (\ref{ff1})-(\ref{ff4}) is well defined in the region
of the phase space where we have:
\begin{equation}\label{ff6}
F(q_1,q_2,\pi_1,\pi_2, x) \neq 0.
\end{equation}
Let us find the fixed points of the vector field
(\ref{ff1})-(\ref{ff4}). Replacing the left-hand side of each
equation by zero leads to
\begin{eqnarray}
\label{f1}  0 &=&  - \pi_1 + \pi_2 +x ,\\
 \label{f2} 0 &=&  \pi_1 - \pi_2 +x,\\
\label{f3} 0 &=&  2 e^{2 q_1}- e^{-q_1 + q_2}  ,\\
\label{f4} 0 &=& - F + e^{-q_1 + q_2}.
\end{eqnarray}
The solution to (\ref{f1}) and (\ref{f2}) is easily found to be
\begin{equation}\label{d1}
    x=0, ~~~\pi_1 - \pi_2 =0 .
\end{equation}
The analysis of (\ref{f3}) and (\ref{f4}), taking into account
(\ref{d1}) and applying the reasoning of Sec. III leads to the
condition
\begin{equation}\label{d2}
\pi_1 = 0 = \pi_2,~~~q_1 \rightarrow - \infty,~~~q_2 - q_1
\rightarrow - \infty .
\end{equation}
However, the condition (\ref{d2}) does not satisfy the condition
(\ref{ff6}). Therefore, the system (\ref{ff1})-(\ref{ff4})
satisfying the condition (\ref{ff6}) is {\it regular}, i.e., does
not have critical points.

Let us examine the consequences of the breaking of (\ref{ff6}).
One can easily find that, in such a case, we have
\begin{equation}\label{d3}
H_T - q_2 = q_3 - q_2  \rightarrow - \infty,~~~0 = T =-\pi_3 = -
p_3 ,
\end{equation}
which means that the breaking of (\ref{ff6}) takes place for the
set of critical points $S_B$ defined by (\ref{critS0}).

One may wonder what the advantage is in using the reduced phase
space Hamiltonian scheme, compared to the scheme prior to the
reduction. It is clear that in the former case  we get the true
Hamiltonian that can be later used in describing the {\it quantum}
evolution of the system. Having a physical Hamiltonian is the
``first things first'' of our quantization scheme
\cite{Malkiewicz:2009qv,Malkiewicz:2010py,Mielczarek:2011mx,
Malkiewicz:2011sr,Mielczarek:2012qs}. In the latter case, an
evolution of the system is defined, at the quantum level, by
rewriting the constraint operator in the form leading to the
Schr\"{o}dinger-like equation (see, e.g., \cite{Ashtekar:2011ni}
for more details). This procedure implements quantum evolution in
a more complicated way than within our approach.

\section{Conclusions}

The novelty of our approach is  an atypical derivation of the
Hamiltonian description of the nondiagonal Bianchi IX dynamics.
Our point of departure is taking the Einstein equations in the
asymptotic region near the cosmological singularity. Next, we make
an educated guess as to what  Lagrangian and Hamiltonian could
lead to these equations by using the principle of least action.
Our approach  guarantees that we are really dealing with the
prototype of the BKL scenario, as the BKL conjecture was derived
by generalization  of the dynamics of the nondiagonal Bianchi IX
model \cite{BKL33}.

Taking into account the Hamiltonian constraint in two ways, before
or after identifying the critical points, leads to the same
results concerning the set of critical points of the vector field
of the considered dynamical system. The situation in which
critical points occur in {\it asymptotic} regions is not an
obstacle because one may project an unbounded subspace onto a
finite region. The original phase space presented in Sec. III is
{\it higher} dimensional. An additional difficulty results from
the fact that the set of critical points is not a set of  isolated
points, but a three-dimensional {\it continuous} subspace of $
\bar{\dR}^6$. The perturbation of the vector field in the
neighborhood of such a space, seems to require  an analysis in the
{\it transverse} space to the space of critical points.

Since {\it all} eigenvalues of the Jacobian (corresponding to the
nonlinear vector field) are purely imaginary, no reduction to
lower-dimensional phase space is possible by using, e.g., the
center manifold theory \cite{Perko, Wiggins}. Since all the
critical points are {\it nonhyperbolic}, the information obtained
from linearization is inconclusive. One has to apply, e.g., the
theory of normal forms \cite{Perko, Wiggins} to get insight into
the structure of the space of orbits in the neighborhoods of the
critical points.  The nonhyperbolicity seems to be a generic
feature of the considered singular dynamics. Making use of the
McGehee (Appendix B) transformations leads to the same type of
criticality of the phase space.

To cope with some of the problems described above we propose a new
approach (presented in Sec. IV).  Our method opens the door for
finding the {\it physical} Hamiltonian  and canonically conjugated
{\it time} variable. Using partial Dirac observables as new
variables in the physical phase space enables an identification of
the Hamiltonian structure. As far as we know, almost all methods
applied so far for the analysis  of the phase space structure of
the Bianchi IX model have used noncanonical transformations of the
phase space and time variables.  If we are only interested in the
{\it classical} dynamics, such a procedure is satisfactory.
However, this is not the case when one plans {\it quantization}:
two classical systems not related by the canonical transformations
become unitarily inequivalent after quantization, i.e., describe
{\it different} quantum systems. On the contrary, our method is
based on {\it canonical} transformations\footnote{Even in the case
when one considers two classical theories related by a canonical
transformation, it is not always the case that the quantum
theories that result from them are unitarily  related.}. Apart
from this, describing an evolution of a {\it quantum} system
requires an identification of the physical Hamiltonian generating
classical dynamics, which after quantization turns into a
self-adjoint operator. Such an operator can be used to define a
unitary operator due to the Stone theorem. The issue of an
evolution of the {\it quantum} Bianchi IX model has not been
considered so far, to such an extent as  has been done for the FRW
and the Bianchi I models  within the reduced phase space
quantization method (see, e.g.,
\cite{Malkiewicz:2009qv,Malkiewicz:2010py,Mielczarek:2011mx,
Malkiewicz:2011sr,Mielczarek:2012qs}).

In both the cases of absolute and relative dynamics, the
determination of the physical  Hamiltonian and time is not unique.
There are many possible choices for these two quantities. This is
connected to the known problem of the choice of time in
gravitational systems, which occurs both at classical and quantum
levels (see, e.g., \cite{Malkiewicz:2011sr} for more details).

In the future we plan to find the true Hamiltonian  (i) after
analytically solving the dynamics using the method of Sec. V and
(ii) within the method of relative dynamics of Sec. VI. In the
former case, we plan to make an extension of the procedure tested
within the first iteration. One may try to find an exact solution
using an induction method applied to the sequence of successive
approximations. In the latter case, we plan to implement the
definition of time that  changes monotonically {\it globally}.
Such a form has the time $T := x_1 + x_2 + x_3$ because $\exp{(x_1
+ x_2 + x_3)} = a b c$, so it is proportional to the local volume
of space that decreases monotonically to zero near the
cosmological singularity (indicating the singularity).

Promising tools to be used within the dynamical system methods
were proposed a long time ago by Bogoyavlensky \cite{Bogo}. Issues
such as higher dimensionality of phase space, criticality of
continuous sets, asymptotic criticality, and, finally
nonhyperbolicity of critical sets do not seem to be problematic in
this approach. Bogoyavlensky's method has been developed by Uggla
and collaborators in a series of papers (see, e.g.,
\cite{Heinzle:2009eh,Heinzle:2009du} and references therein).
Their Hubble-normalized bounded variables seem to cope with the
problem of fixed points ``at infinity'' of ``non-hyperbolic''
type. This method concerns, however, the {\it diagonal} Bianchi IX
model. Trying to adopt this method to our model is another aspect
of our program. An issue ofv relevance for us, while applying the
sophisticated variables on phase space proposed in Refs.
\cite{Bogo,Heinzle:2009eh}, is that of  {\it noncanonical}
transformations on phase space which are used to implement them.
We present some aspects of this problem in Appendix B, where we
apply McGehee-type transformations. However, a complete discussion
of this problem is beyond the scope of the present paper and will
be given elsewhere.

\acknowledgments

It is a pleasure to thank Vladimir Belinski for many valuable
discussions on the BKL scenario, which has inspired us to begin
working on the dynamics of the Bianchi IX model. We are grateful
to the ICRANet for financial support during our visits to this
institute.  We also thank   Claes Uggla and  Stephen Wiggins  for
the suggestions concerning  the dynamical systems method, and
Przemys{\l}aw Ma{\l}kiewicz for valuable discussions concerning
the factorization method and the problem of time. Finally, we are
very grateful to an anonymous referee for inspiring questions,
suggestions, and remarks.

\appendix

\section{Reduced 2-form }
An imposition of the constraint (\ref{x30}) onto (\ref{facHH})
leads to the expression

\begin{eqnarray}\label{mega}
\Omega &=&
\Big(-\frac{1}{2} e^{-{x_{10}}} \left(2 e^{3 {x_{10}}} {p_{10}}-2 e^{{x_{20}}} {p_{10}}
-4 e^{3 {x_{10}}} {p_{20}}+3 e^{{x_{20}}} {p_{20}}\right)   d{x_{10}}\nonumber \\
&& -\frac{1}{2} e^{-x_{10}+x_{20}} (2 p_{10}-3 p_{20}) dx_{20}+ \frac{1}{2} e^{-x_{10}}
\left(e^{3 x_{10}}+2 e^{x_{20}}-e^{x_{10}} p_{20}^2\right) dp_{10} \nonumber\\
&& - \frac{1}{2} e^{-x_{10}} \left(2 e^{3 x_{10}}+3
e^{x_{20}}-e^{x_{10}} p_{10}^2+2 e^{x_{10}} p_{10} p_{20}-2
e^{x_{10}} p_{20}^2\right) dp_{20}\Big)\wedge d{T}\nonumber
 \\
&&+\Big(-\left(e^{3 x_{10}}+e^{x_{20}}-e^{x_{10}} p_{10}
p_{20}+e^{x_{10}} p_{20}^2\right)^{-1}e^{-x_{10}} \left(-2 e^{4
x_{10}}+e^{x_{10}+x_{20}}+4 e^{6 x_{10}} {T}-3 e^{2 x_{20}}
{T}\right.\nonumber
 \\
&&\left. +e^{3 x_{10}} +x_{20} {T}-4 e^{4 x_{10}} p_{10} p_{20}
{T}+3 e^{x_{10}+x_{20}} p_{10} p_{20} {T}+4 e^{4 x_{10}} p_{20}^2
{T}-3 e^{x_{10}+x_{20}} p_{20}^2 {T}\right) dp_{20}\nonumber
 \\
&&
+\left(e^{3 x_{10}}+e^{x_{20}}-e^{x_{10}} p_{10} p_{20}+e^{x_{10}} p_{20}^2\right)^{-1}e^{-x_{10}}
\left(-3 e^{4 x_{10}}+e^{2 x_{10}} p_{10} p_{20}-e^{2 x_{10}} p_{20}^2+2 e^{6 x_{10}} {T}\right.\nonumber\\
&& \left.-2 e^{2 x_{20}} {T}-2 e^{4 x_{10}} p_{10} p_{20} {T}+2
e^{x_{10}+x_{20}} p_{10} p_{20} {T}+2 e^{4 x_{10}} p_{20}^2 {T}-2
e^{x_{10}+x_{20}} p_{20}^2 {T}\right) dp_{10}\Big)\wedge
dx_{10}\nonumber
\\
&& +\Big(-e^{-x_{10}+x_{20}} (-e^{x_{10}}+3 e^{3 x_{10}} {T}+3
e^{x_{20}} {T} -3 e^{x_{10}} p_{10} p_{20} {T}+3 e^{x_{10}}
p_{20}^2 {T}) \nonumber
\\
&& \left(e^{3 x_{10}}+e^{x_{20}}-e^{x_{10}} p_{10}
p_{20}+e^{x_{10}} p_{20}^2\right)^{-1}dp_{20}\nonumber
\\
&&
-e^{-x_{10}} \left(e^{4 x_{10}}+2 e^{x_{10}+x_{20}}-e^{2 x_{10}} p_{10} p_{20}+e^{2 x_{10}}
 p_{20}^2-2 e^{2 x_{20}} {T}-2 e^{3 x_{10}+x_{20}} {T} \right.\nonumber
\\
&& \left.+2 e^{x_{10}+x_{20}} p_{10} p_{20} {T} -2
e^{x_{10}+x_{20}} p_{20}^2 {T}\right) \nonumber
\\
&& \left(e^{3 x_{10}}+e^{x_{20}}-e^{x_{10}} p_{10}
p_{20}+e^{x_{10}} p_{20}^2\right)^{-1}dp_{10}\Big)\wedge dx_{20}
\end{eqnarray}

\section{McGehee-type variables}

In this section, we apply the McGehee-type  transformations
\cite{RMG}, which have been successfully  applied to the analysis
of dynamical systems of general relativity (see, e.g.,
\cite{DeOliveira:2002ih,Belbruno:2011nn}).

We define the transformation
\begin{equation}\label{m1}
\dR^6 \ni (x_1,x_2,x_3,p_1,p_2,p_3)\longrightarrow (u_1,u_2,u_3,
v_1,v_2,v_3) \in \dR^3_+ \times \dR^3
\end{equation}
as follows:
\begin{equation}\label{m2}
(u_1,u_2,u_3, v_1,v_2,v_3):= (\exp x_1,\exp x_2,\exp x_3, p_1,
p_2, p_3).
\end{equation}

We would like to interpret the dynamics in the new coordinates in
terms of our original coordinates. This is possible if the mapping
is {\it canonical} \cite{HG,AM}:
\begin{equation}\label{canmc1}
\{u_k,u_l\}_{x,p} = 0 = \{v_k,v_l\}_{x,p},~~~~\{u_k,~v_l\}_{x,p} =
\delta_{kl},
\end{equation}
where
\begin{equation}\label{poiss}
\{\cdot,\cdot\}_{x,p}:= \sum_{k=1}^3
\Big(\frac{\partial\cdot}{\partial
x_k}\frac{\partial\cdot}{\partial
p_k}-\frac{\partial\cdot}{\partial
p_k}\frac{\partial\cdot}{\partial x_k}\Big)
\end{equation}
One may easily verify that
\begin{equation}\label{canmc2}
\{u_k,u_l\}_{x,p} = 0 = \{v_k,v_l\}_{x,p},~~~~\{u_k,~v_l\}_{x,p} =
u_k \delta_{kl}.
\end{equation}
Therefore, the transformation (\ref{m1}) is not canonical. Thus,
the phase space in the new variables corresponds to a {\it
different} dynamical system; i.e., the transformation maps the
phase space of our original system to the phase space of a new
system. As long as we stay at the classical level, i.e. we do not
plan on making quantization of our classical system, such a
procedure may be reasonable.  This is true if the new system turns
out to have simpler dynamics than the original one. In the rest of
the present section we verify this expectation.

Under the transformation (\ref{m1}), the vector field
(\ref{x1})-(\ref{p3}) turns into
\begin{eqnarray}
2\dot{u}_1 &=& u_1 (-v_1 +v_2 + v_3), \label{cm1}\\
2\dot{u}_2 &=& u_2 (v_1 -v_2 + v_3), \label{cm2}\\
2\dot{u}_3 &=& u_3 (v_1 +v_2 - v_3), \label{cm3}\\
\dot{v}_1 &=& 2 u_1^2 - u_2/u_1, \label{cm4}\\
\dot{v}_2 &=& u_2/u_1 - u_3/u_2 , \label{cm5}\\
\dot{v}_3 &=& u_3/u_2 . \label{cm6}
\end{eqnarray}
To have a well-defined vector field,  possibly of class $C^r$, we
rescale the time variable as follows:
\begin{equation}\label{tres}
\dot{f}:= \frac{df}{d\tau}=\frac{df}{ds}\frac{ds}{d\tau}:=
\frac{1}{u_1 u_2}\frac{df}{ds}=: \frac{1}{u_1 u_2}f'.
\end{equation}
Making use of (\ref{tres}) turns (\ref{cm1})-(\ref{cm6}) into
\begin{eqnarray}
2{u}_1' &=& u_1^2 u_2 (-v_1 +v_2 + v_3), \label{mc1}\\
2{u}_2' &=& u_1 u_2^2 (v_1 -v_2 + v_3), \label{mc2}\\
2{u}_3' &=& u_1 u_2 u_3 (v_1 +v_2 - v_3), \label{mc3}\\
{v}_1' &=& 2 u_1^3 u_2 - u_2^2, \label{mc4}\\
{v}_2' &=& u_2^2 - u_1 u_3 , \label{mc5}\\
{v}_3' &=& u_1 u_3. \label{mc6}
\end{eqnarray}
We also redefine the constraint  (\ref{H33}) by  multiplying it by
the factor $u_1 u_2$. The new form of the constraint reads:
\begin{equation}\label{newc}
0 = u_1 u_2 H =: K,
\end{equation}
where
\begin{equation}\label{conmc}
K = \frac{1}{2}u_1 u_2 (v_1 v_2 + v_1 v_3 + v_2 v_3) -
\frac{1}{4}u_1 u_2 (v_1^2 + v_2^2 + v_3^2)- u_1^3 u_2 - u_2^2 -u_1
u_3.
\end{equation}

All considerations carried out so far have been done with the
assumption that $u_1 > 0, u_2 > 0$ and $u_3 > 0$. From now on, we
change this assumption; i.e., we may have $u_1 \geq 0,~ u_2 \geq
0$, and $u_3 \geq 0$. One can say that we extend the phase space
of our system to a bigger one.

Let us identify the critical points of the vector field
(\ref{mc1})-(\ref{mc6}). We find that the set of  critical points
is defined to be
\begin{equation}\label{critmc}
S_M:= \{(u_1,u_2,u_3, v_1,v_2,v_3)\in \dR^6 ~|~u_1 = 0 = u_2,
u_3\geq 0, (v_1,v_2,v_3)\in \dR^3\}.
\end{equation}
It is clear that any element of $S_M$ satisfies the constraint
(\ref{conmc}). One may verify that the transformation (\ref{m1})
does not map the critical subspace $S_B$ into $S_M$.

Let us examine the type of criticality specific to the set of
critical points (\ref{critmc}). It is not difficult to find that
the Jacobian of (\ref{mc1})-(\ref{mc6}) at any point of the
critical set $S_M$ reads:
\begin{equation}\label{mcJ}
\nonumber J=\left(\begin{array}{cccccc}
    0 & 0  & 0  & 0 &  0 &  0 \\
    0 & 0  & 0  &  0 & 0 &  0 \\
    0 & 0  & 0  & 0  & 0  & 0 \\
    0 & 0  & 0  & 0  & 0  & 0 \\
 -u_3 & 0  & 0 & 0   & 0  & 0 \\
  u_3 & 0 & 0  & 0   & 0  & 0
\end{array} \right)
\end{equation}
The characteristic polynomial associated with $J$ reads $
P(\lambda)=\lambda^6$.  Thus, $\lambda =0$ is an eigenvalue of $J$
of multiplicity 6. Therefore, the set $S_M$ consists of
nonisolated {\it nonhyperbolic} fixed points. This means that
linearization of the exact vector field, at the set of critical
points $S_M$, cannot help us understand the mathematical structure
of the space of orbits of the considered vector field. An
examination of the nonlinearity cannot be avoided.

\end{document}